\documentclass[superscriptaddress,aps,prl,reprint]{revtex4-1}

\newcommand{\bra}[1]{\langle #1 |}			
\newcommand{\ket}[1]{| #1 \rangle}

\renewcommand{\Re}{\operatorname{Re}}
\renewcommand{\Im}{\operatorname{Im}}

\usepackage{commath}

\usepackage[table]{xcolor}
\usepackage{amsmath,amssymb}
\usepackage[utf8]{inputenc}
\usepackage{graphicx}
\usepackage[caption=false,subrefformat=parens,labelformat=parens]{subfig}
\usepackage{epstopdf}
\usepackage{siunitx}
\usepackage{tikz}
\usepackage{placeins}
\usepackage{cleveref}

\begin{document}
\begin{abstract}
Topological insulator films are promising materials for optoelectronics due to a strong optical absorption and a thickness dependent band gap of the topological surface states. They are superior candidates for photodetector applications in the THz-infrared spectrum, with a potential performance higher than graphene. Using a first-principles $k\cdot p$ Hamiltonian, incorporating all symmetry-allowed terms to second order in the wave vector $k$, first order in the strain $\epsilon$ and of order $\epsilon k$, we demonstrate significantly improved optoelectronic performance due to strain. For Bi$_2$Se$_3$ films of variable thickness, the surface state band gap, and thereby the optical absorption, can be effectively tuned by application of uniaxial strain, $\epsilon_{zz}$, leading to a divergent band edge absorbance for $\epsilon_{zz}\gtrsim 6\%$. Shear strain breaks the crystal symmetry and leads to an absorbance varying significantly with polarization direction. Remarkably, the directional average of the absorbance always increases with strain, independent of material parameters.
\end{abstract}

\title{Strain-enhanced optical absorbance of topological insulator films}

\author{Mathias Rosdahl Brems}
\affiliation{Technical University of Denmark, Department of Photonics Engineering, Kgs. Lyngby, 2800, Denmark}

\author{Jens Paaske}
\affiliation{Center for Quantum Devices, Niels Bohr Institute, University of Copenhagen,
Universitetsparken 5, DK-2100 Copenhagen, Denmark}

\author{Anders Mathias Lunde}
\affiliation{Center for Quantum Devices, Niels Bohr Institute, University of Copenhagen,
Universitetsparken 5, DK-2100 Copenhagen, Denmark}

\author{Morten Willatzen}
\affiliation{Technical University of Denmark, Department of Photonics Engineering, Kgs. Lyngby, 2800, Denmark}

\maketitle

Three-dimensional topological insulators (TI) like Bi$_2$Se$_3$, Bi$_2$Te$_3$ and Sb$_2$Te$_3$ have relatively large inverted band gaps of the order of 0.3 eV and thus host robust topological surface states (TSS)~\cite{oldmodel,model} whose Dirac cone band structures have been studied extensively using angle resolved photoemission spectroscopy~\cite{experiment_gap,experiment_bi2te3,experiment_bi2se3,experimental_gap_osc}. These are layered materials with van der Waals bonded quintuple layers (QL) and can be exfoliated to produce thin TI films for which TSS on opposite surfaces overlap and gap out the respective Dirac cones. Already for a 6 QL film, these Dirac gaps are unobservably small and the film effectively reverts to ordinary 3D-TI behavior~\cite{Zhang2010}.

Zhang et al.~\cite{Zhang_optical} pointed out that TI films should be promising materials for optoelectronics applications, with an optical absorbance due to the TSS comparable to that of graphene ($\pi\alpha\approx 2.3\%$ given in terms of the fine structure constant $\alpha=e^2/(4\pi\epsilon_{0}\hbar c)\approx 1/137$), roughly two orders of magnitude higher than  conventional photodetector materials like Hg$_{1-x}$Cd$_x$Te. One surface of a TI film hosts only a single gapless Dirac cone, which gives rise to a frequency independent total absorbance of $\pi\alpha/2$. For thin ($\lesssim$ 6 QL) TI films, the gapped Dirac electrons may enhance this up to $\pi\alpha$ and even larger, for frequencies close to the TSS band gap.
Breaking inversion symmetry of a thin TI film shifts the minimum Dirac gap to finite wave vectors~\cite{njp_films} and a Mexican-hat shape of the surface state conduction band gives rise to a Van Hove singularity in the joint density of states (JDOS), leading to a divergent band edge absorption and tunability of the two-photon absorption spectrum~\cite{Wang_optical}. Furthermore, the hexagonal warping in the TSS bandstructure has been shown to enhance the optical absorbance above the TSS band gap relevant for higher frequencies~\cite{Shao2014}. Recent experimental investigations of photocurrent response have indeed shown promising optoelectronic functionality for different TI based nano, and hybrid structures~\cite{Zang_bi2se3nanosheets,Sharma_nanowires,Zheng_photodetectorSb2Se3,Zhang_polycrystallinefilm,Qiao_grapheneTI,Giorgianni2016}.


In this paper, we investigate the effects of strain on the optical absorption of Bi$_2$Se$_3$ films of variable thickness. We employ the most general Hamiltonian for the Bi$_2$Se$_3$ class of materials, including all symmetry-allowed terms to second order in the wave vector $k$, first order in the strain tensor $\epsilon$ and of order $\epsilon k$. The detailed derivation of this $k\cdot p$ Hamiltonian and its surface states, using the method of invariants~\cite{LewYanVoon}, will be presented in a separate publication~\cite{mypaper}. Previous theoretical studies of strain on  Bi$_2$Se$_3$ were based on density functional theory~\cite{luo,young}, and have focused only on the change of the bulk band gap and the possibility of a strain-induced topological phase transition. In contrast, our model Hamiltonian provides deeper insight into the electronic structure and, in particular, how the surface states are affected by strain. To describe the low-energy physics of a thin film, we derive an effective 2D model for the surface states, thus disregarding contributions from the bulk states, which at photon energies in the THz-infrared spectrum ($\omega~\sim$~30-300 meV) are expected to be largely negligible~\cite{Li2015}.

Using this surface-state model together with Fermi's Golden Rule, we determine the optical absorbance in the presence of strain. Basically, we find that uniaxial strain decreases the bulk band gap and thereby increases the spatial extent of the TSS wave functions, which in turn leads to a larger inter-surface overlap and concomitant energy gap of the TSS. In this way, straining a TI film is perceived by the surface states as making the film thinner, thus enhancing the band-edge absorption studied in Ref.~\onlinecite{Zhang_optical}. The absorption edge moves to higher frequency, and the absorbance at the edge increases. Increasing the uniaxial strain above $6$~\%, we find that the minimum TSS band-gap shifts to finite wave vectors. This leads to a Van Hove singularity in the joint density of states (JDOS), resulting in a divergent band edge absorption even in the presence of inversion symmetry. For shear strain or $\epsilon_{xx}\neq\epsilon_{yy}$ the isotropy of the model is broken, and a strong polarization dependence appears. In an isotropic model the photocurrent is polarization dependent due to the anisotropic excitation in $k$ space~\cite{Yao2015}, and we propose to enhance this effect by strain. Subject to strain, in the thick-film limit where the TSS band gap closes, the absorbance is independent of the photon energy, with a magnitude depending sinusoidally on the polarization angle. Averaging over polarization angle, however, we show that strain always increases the average absorbance above the universal value $\frac{\alpha\pi}{2}$, independently of material parameters.

{\it Model Hamiltonian.}
Based on the symmetries of the crystal, we have employed the Method of Invariants to derive the most general $k\cdot p$ Hamiltonian to second order in wave vector and first order in strain~\cite{mypaper}, given here in the basis~\cite{model}  $\ket{P1_-^+,\tfrac{1}{2}}$, $-i\ket{P2_+^-,\tfrac{1}{2}}$, $\ket{P1_-^+,-\tfrac{1}{2}}$, $i\ket{P2_+^-,-\tfrac{1}{2}}$ by
\begin{align}\label{eq:bhzH}
H^{3D} = \mathcal E_0 &+ \begin{pmatrix}
\mathcal M & \beta^*_i k_i & 0 & \alpha^*_i k_i \\
\beta_i k_i & -\mathcal M & \alpha^*_i k_i & 0 \\
0 & \alpha_i k_i & \mathcal M & - \beta_i k_i \\
\alpha_i k_i & 0 & -\beta^*_i k_i & -\mathcal M
\end{pmatrix}
\end{align}
where repeated indices are summed over $i=x,y,z$, and
\begin{align}
 \mathcal E_0(\mathbf k) &=  C + C_{1} \epsilon_{zz} + C_{2}  \epsilon_{\|} + D_1 k_z^2 + D_2 k_{\|}^2, \\
\mathcal M(\mathbf k) &= M + M_1 \epsilon_{zz} + M_2 \epsilon_{\|}- B_1 k_z^2 - B_2 k_{\|}^2, \\
\alpha_x &=  A_2 + A_{21}\epsilon_{zz} + A_{22}\epsilon_{\|}  + i Y_3 \epsilon_{z-} + Y_4\epsilon_+,\\
\alpha_y &= i A_2 + iA_{21}\epsilon_{zz} + iA_{22}\epsilon_{\|} +  Y_3 \epsilon_{z-} - i Y_4\epsilon_+, \\
\alpha_z &= Y_1\epsilon_{z+} + i Y_2 \epsilon_-,\\
\beta_x &= X_1 \epsilon_{zx} +  2X_2\epsilon_{xy} - iX_3  \epsilon_{zy} + iX_4(\epsilon_{xx}-\epsilon_{yy}),\\
\beta_y &= X_1\epsilon_{zy} + X_2(\epsilon_{xx}-\epsilon_{yy}) + i X_3 \epsilon_{zx} - 2i X_4 \epsilon_{xy}, \\
\beta_z &=  A_1 + A_{11}\epsilon_{zz} + A_{12}\epsilon_{\|},
\end{align}
with $\epsilon_{\|} = \epsilon_{xx} +\epsilon_{yy}$, $\epsilon_{\pm} = \epsilon_{xx} - \epsilon_{yy} \pm 2i\epsilon_{xy}$, $\epsilon_{z\pm} = \epsilon_{zx} \pm \epsilon_{zy}$, and ${\bf k} = ({\bf k}_{\|},k_z)$. All capital roman letters denote real parameters of the model, which are not determined by the symmetries of the crystal.

\begin{figure}[t]
\includegraphics[width=\columnwidth]{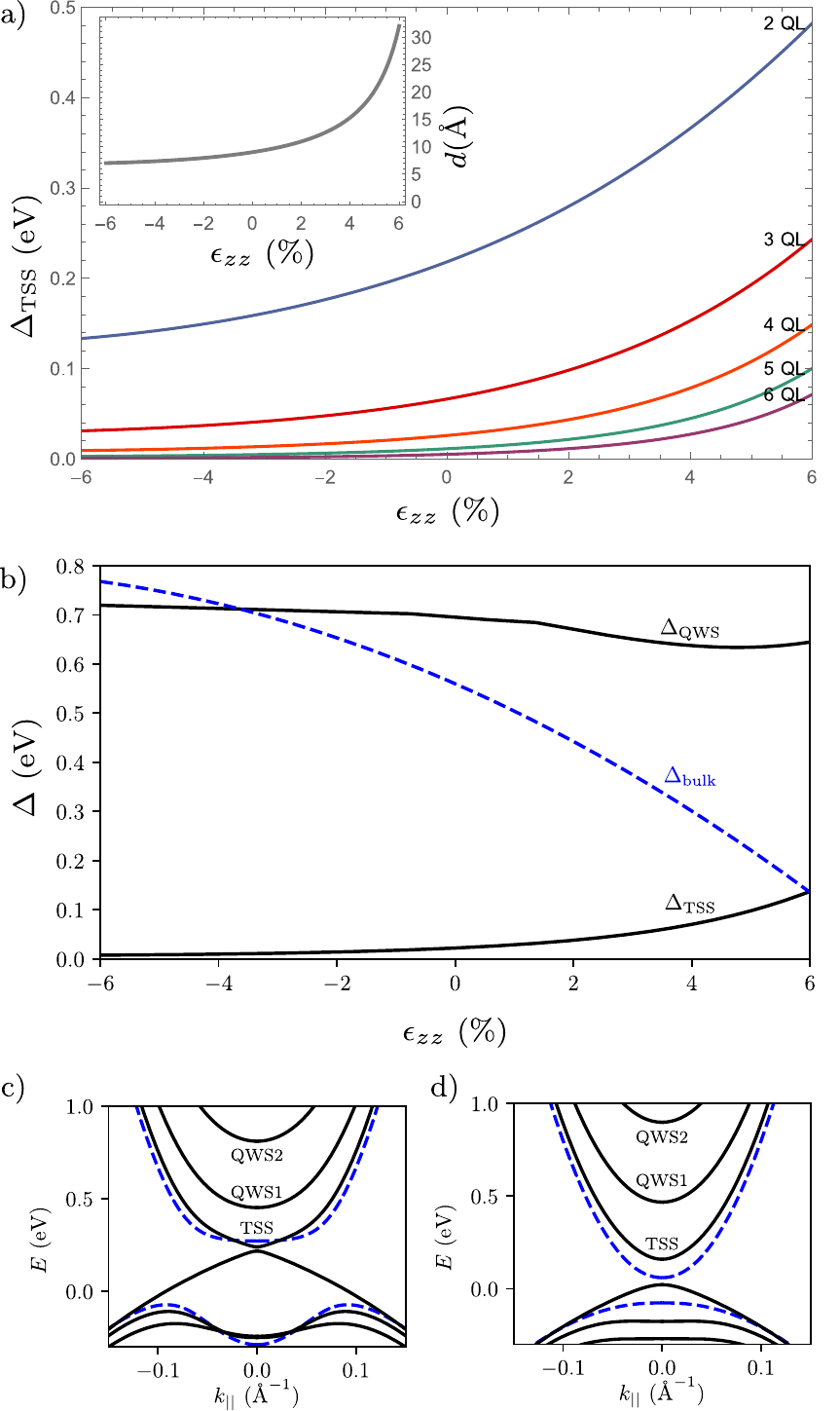}
\caption{In the upper plot, we show the surface state band gap $\Delta$ calculated from $H^{3D}$ for a slab geometry with $2-6$ QL, using the parameters from Ref.~\onlinecite{model} and the dependence of the bulk band gap with strain from Ref. \onlinecite{young}. The inset shows the expectation value of the distance $d$ to the surface of a semi-infinite slab vs. $\epsilon_{zz}$ for a surface state electron with $k_{\|} = 0$. As the surface state extends further into the bulk, the hybridization gap increases. In the middle panel, we plot the band gaps at $k_{\parallel} = 0$ for a $4$QL film associated with surface states (TSS), quantum-well states (QWS), and, for comparison, also the band gap of bulk Bi$_2$Se$_3$ computed with DFT~\cite{young}. In the left and right plots of the lower panel, the TSS and QWS bands of a $4$QL film are shown for $\epsilon_{zz} = 0$ and $\epsilon_{zz} = 6$~\%, respectively. For comparison, also the bulk bands are plotted (blue dashed).}
\label{fig:gap}
\end{figure}

For an infinite slab in the $xy$ plane, with a finite thickness $L$ in the $z$ direction, we impose hard-wall boundary conditions at the surfaces located at $\pm L/2$. The full Hamiltonian can be split into two parts  $H^{3D}(\mathbf k)=H_0(k_z) + \Delta H(\mathbf k)$, with $H_0(k_z)$ containing all constant terms, $k_z$ up to second order and first order strain terms and $\Delta H(\mathbf k)$ containing terms first and second order in $\mathbf k_{||}$ and of order $\epsilon \mathbf k$. Treating $\Delta H(\mathbf k)$ as a perturbation, we first solve $H_0(k_z)$ giving the eigenstates ~\cite{njp_films}
\begin{align}
\Psi^\uparrow_\pm(z)=\left(\phi_\pm (z),0\right), \quad
\Psi^\downarrow_\pm(z)=\left(0,\tau_z\phi_\pm (z)\right),
\end{align}
where $\tau_i$ denotes the Pauli matrices and
\begin{align}
\phi_{\pm}(z)&=C_{\pm}
\left((D_1+B_1)\eta_{\pm}\varphi_{\pm}(z), iA_1 \varphi_{\mp}(z)\right).
\end{align}
Here $C_\pm$ are normalization constants, and
\begin{align}
\eta_{\pm}&=\frac{\lambda_2^2-\lambda_1^2}{
\lambda_{1}\varrho_{\mp}(\frac{\lambda_1 L}{2})/\varrho_{\pm}(\frac{\lambda_1 L}{2})-
\lambda_{2}\varrho_{\mp}(\frac{\lambda_2 L}{2})/\varrho_{\pm}(\frac{\lambda_2 L}{2}) },\nonumber\\
\varphi_{\pm}(z) &=\left( \frac{\varrho_{\pm}(\lambda_1 z)}{\varrho_{\pm}(\frac{\lambda_1 L }{2})} - \frac{\varrho_{\pm}(\lambda_2 z)}{\varrho_{\pm}(\frac{\lambda_2 L }{2}) }\right),
\end{align}
with $\varrho_{\pm}(x)=(e^{x}\pm e^{-x})/2$ encoding hyperbolic sines and cosines, and
$\lambda_{\alpha}=(((-1)^\alpha\sqrt{R}-F)/(2D_{+}D_{-}))^{1/2}$, where
\begin{align}
F&=|A_1|^{2} + 2((E - \mathcal E_{\mathbf k=0})D_{1} - \mathcal M_{\mathbf k=0} B_{1}),\\
R&=F^2+4D_{+}D_{-}(\mathcal M_{\mathbf k=0}^{2}-(E-\mathcal E_{\mathbf k=0})^{2}),
\end{align}
with $D_{\pm}=D_{1}\pm B_{1}$. To obtain an effective 2D model for the surface states we project $H^{3D}(\mathbf k)$ into the surface state basis $\left\{\ket{\Psi_-^\uparrow},\ket{\Psi_+^\downarrow},\ket{\Psi_+^\uparrow},\ket{\Psi_-^\downarrow}\right\}$ of $H_0(k_z)$, thus excluding contributions from the bulk states:
\begin{eqnarray}
\lefteqn{H^{2D} = \bar E_0 + \bar D k_{\|}^2} \\
&&+\begin{pmatrix} \bar M + \bar B k_{\|}^2 & \bar \alpha_i^* k_i & \bar \beta_i k_i & 0\\
\bar \alpha_i k_i & -\bar M - \bar B k_{\|}^2 & 0 & \bar \beta_i k_i \\
\bar \beta^*_i k_i & 0 & -\bar M - \bar B k_{\|}^2 & - \bar \alpha_i^* k_i \\
0 & \bar \beta^*_i k_i & -\bar \alpha_i k_i & \bar M + \bar B k_{\|}^2
\end{pmatrix},\nonumber
\end{eqnarray}
where repeated indices are summed over $i=x,y$. The parameters of the 2D model, denoted with a bar, now depend on thickness, $L$, and are related to the strain-dependent parameters of the 3D model by:
\begin{align}
\bar E_0 &= \frac{E^{0}_+ + E^0_-}{2},\\
\bar M &= \frac{E^0_- - E^0_+}{2},\\
\bar D &= D_2 + \frac{\bra{\phi_+}\tau_z\ket{\phi_+} + \bra{\phi_-}\tau_z\ket{\phi_-}}{2}B_2,\\
\bar B &= \frac{\bra{\phi_+}\tau_z\ket{\phi_+} - \bra{\phi_-}\tau_z\ket{\phi_-}}{2}B_2 ,\\
\bar \alpha_i &= i\bra{\phi_+}\tau_y\ket{\phi_-} \alpha_i ,\\
\bar \beta_i &=\bra{\phi_+}\tau_x\ket{\phi_-}\Re[\beta_i] +  i\bra{\phi_+}\tau_y\ket{\phi_-} i\Im[\beta_i],
\end{align}
\noindent where $E^0_\pm$ are the eigenenergies of $H_0(k_z)$. From  this, we obtain an effective 2D model and derive the spectrum for TSS on a strained TI film:
\begin{align}
E_\pm({\mathbf k}_{\|})={\bar E}_{0}+{\bar D}k_{\|}^{2}\pm\frac{1}{2}\Delta E_{{\mathbf k}_{\|}},
\end{align}
with $\Delta E_{{\mathbf k}_{\|}} = 2 \sqrt{|\bar \alpha_i k_i|^2 + |\bar \beta_i k_i|^2 + (\bar M + \bar B k_{\|}^2)^2}$. From the latter result, we obtain the surface state band gap at $k_{\|} = 0$
\begin{align}
 \Delta & = 2\bar M,
\end{align}
which is plotted in Fig.~\ref{fig:gap} as a function of the strain component $\epsilon_{zz}$ for different numbers of quintuple layers in the film. We note in passing that the bulk band gap $\Delta_{bulk}$ of strained Bi$_2$Se$_3$ is given by
\begin{align}
 \Delta_{bulk} & = 2 \left( M + M_1 \epsilon_{zz} + M_2 \epsilon_{\|}\right).
\end{align}
In Fig.~\ref{fig:gap}, we also plot, for a $4$QL film, the band gaps and bandstructure associated with surface states (TSS) and quantum-well states (QWS). For comparison the bulk band gap and bandstructure are shown as blue dashed lines.

{\it Optical absorbance.}
Within this strained 2D model for the TSS of a thin film, we now calculate the optical absorbance using Fermi's Golden Rule in the dipole approximation. Here we consider only linearly polarized, normal incident light, corresponding to the vector potential $ \mathbf A(t) = A\hat{\bf n} \cos(\omega t)$ at the surface, where $\hat{\bf n} =(\cos(\theta_A),\sin(\theta_A),0)$ denotes the polarization vector, given in terms of its angle with the $x$-axis, $\theta_A$. The interaction with the TSS is obtained by minimal substitution, $\mathbf k \rightarrow \mathbf k + e \mathbf A/\hbar$, in $H^{2D}$, which yields the interaction term
\begin{align}
H_\text{int} = \frac{e}{\hbar}
\begin{pmatrix}
0 & \bar \alpha_i^* A_i & \bar \beta_i A_i & 0 \\
\bar \alpha_i A_i & 0 & 0 & \bar\beta_i A_i \\
\bar\beta^*_i A_i & 0 & 0 & -\bar\alpha_i^* A_i  \\
0 & \bar\beta^*_i A_i & -\bar\alpha_i A_i& 0
\end{pmatrix}.
\end{align}
The Golden Rule rate for direct transitions between negative and positive energy surface states is given by
\begin{align}
\Gamma &= \frac{2\pi}{\hbar}\sum_{{\mathbf k}_{\|},i,f } |\bra{f} H_\text{int}\ket{i}|^2
\delta(\hbar\omega-\Delta E_{{\mathbf k}_{\|}}),
\end{align}
and the absorbance, $\mathcal P$, is found as the ratio of absorbed, $\Gamma\hbar\omega/ \text{Area}$, to incident intensity, $\frac{1}{2}c\epsilon_0 \omega^2 |A|^2$, whereby
\begin{multline}
\mathcal P(\theta_A)=
\frac{ \alpha}{2 }\int_0^{2\pi} \mathrm d\theta_k \sum_{k_0(\theta_k)}\\
  \frac{\Im[\chi]^2+|\gamma|^2 +  (\frac{2}{\hbar \omega})^2(\bar M +\bar B k_0(\theta_k)^2)^2 \Re[\chi]^2 }{ A(\theta_k)^2 | A(\theta_k)^2 + 2 \bar B  (\bar M + \bar B k_0(\theta_k)^2)|}.
  \label{absorbance_general}
\end{multline}
Here the sum runs over solutions to $\Delta E_{{\mathbf k}_{\|}}=\hbar\omega$, with the parametrization ${\mathbf k}_{\|}=k_{0}(\cos\theta_{k},\sin\theta_{k})$, given by
\begin{align}
k_0^2(\theta_k)=&\frac{1}{2\bar B^2}\Big(-2\bar M \bar B_2 - A(\theta_k)^2\\
 &\pm \sqrt{(2\bar M \bar B + A(\theta_k)^2)^2 + \bar B^2(\hbar^2\omega^2-4\bar M^2)}\Big),\nonumber
\end{align}
with $A(\theta_k)=\sqrt{|f_{\alpha}(\theta_k)|^2+|f_{\beta}(\theta_k)|^2}$, defined in terms of functions $f_{\alpha}(\theta)=\bar\alpha_x\cos(\theta)+\bar\alpha_y\sin(\theta)$ and
$f_{\beta}(\theta)=\bar\beta_x\cos(\theta)+\bar\beta_y\sin(\theta)$, and with $\chi=f_{\alpha}(\theta_k)f^{\ast}_{\alpha}(\theta_A)+f_{\beta}(\theta_k)f^{\ast}_{\beta}(\theta_A)$ and $\gamma=(\bar \alpha_x \bar \beta_y - \bar \alpha_y \bar \beta_x)\sin(\theta_A - \theta_k)$.

\begin{figure}
\center
\includegraphics[width=\columnwidth]{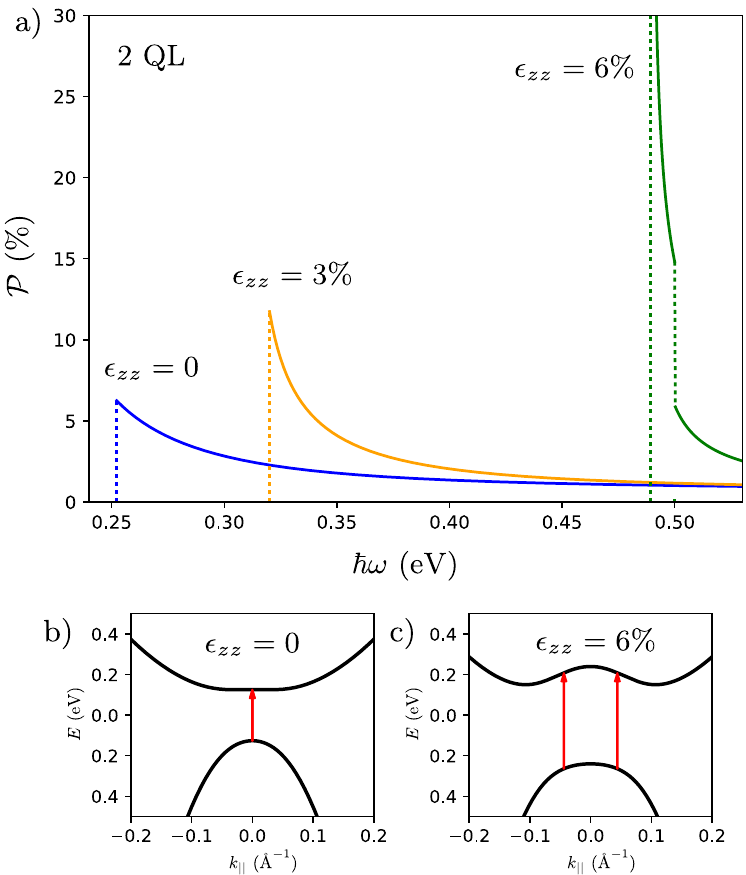} \\
\caption{Upper panel: The absorbance for a $2$~QL slab under tensile uniaxial strain. The absorbance is independent of the light polarization since the isotropy is not broken. The band gap and the band edge absorbance increase with tensile strain. For $\epsilon_{zz}=6\%$ we find a qualitatively different behavior with a band gap at a finite wave vector. This band gap occurs on a circle in $k_x,k_y$ space leading to a diverging band edge absorbance. At the $k_{\|}=0$ band gap we see a discontinuous decrease in the absorbance. Lower panel: $2$QL TSS bands for (left) $\epsilon_{zz} = 0$ and (right) $\epsilon_{zz} = 6$~\%, respectively. The Mexican-hat shape of the TSS conduction band for $\epsilon_{zz} = 6$~\% leads to a Van Hove singularity in the absorbance. We use parameters from Ref.~\onlinecite{njp_films}, and  for $\epsilon_{zz}=3\%$, $6\%$ we use the band gap from the data in  Fig.~\ref{fig:gap}.}
\label{fig:specialcase}
\end{figure}

In the case where only the strain components $\epsilon_{zz}$ and $\epsilon_{\|}$ are non-zero, the rotational symmetry in the $xy$ plane is not broken and Eq.~\eqref{absorbance_general} can be integrated analytically. If furthermore the smallest band gap occurs at $k_{\|}=0$, the result from Ref.~\onlinecite{Zhang_optical} is obtained. However, for  $\bar M \bar B +  A^2(\theta_k) < 0$, the smallest band gap appears at $k_{\|}\neq 0$, which gives rise to a Van Hove singularity in the JDOS causing a divergent absorption edge below the larger band gap at $k_{\|}=0$. As we saw in the previous section, the $k_{\|}=0$ band gap can be significantly increased by application of a tensile uniaxial strain. In Fig.~\ref{fig:specialcase} we show the absorption spectrum for $\epsilon_{zz}=0\%$, $\epsilon_{zz}=3\%$, and $\epsilon_{zz}=6\%$ for a slab thickness of $2$~QL. From 3\% to 6\% we see the band gap and band-edge absorption increasing. For a strain higher than 6\% we find a large absorbance in the energy range between the smallest band gap occurring at $k_{\|}\neq0$ and the $k_{\|}=0$ band gap with a divergence at the former and a discontinuity at the latter. For other thicknesses the change in the band-edge absorbance is similar, but the qualitative change due to the gap occurring at finite wave vector is only observed for 2 QL within a reasonable amount of strain. Also shown in Fig.~\ref{fig:specialcase} are the $2$QL TSS bandstructure for $\epsilon_{zz}=0$ and $\epsilon_{zz}=6$~\%, respectively.

\begin{figure}
\begin{tikzpicture}
\node[inner sep=0pt] (russell) at (0,0)
    {\includegraphics[width=0.5\textwidth]{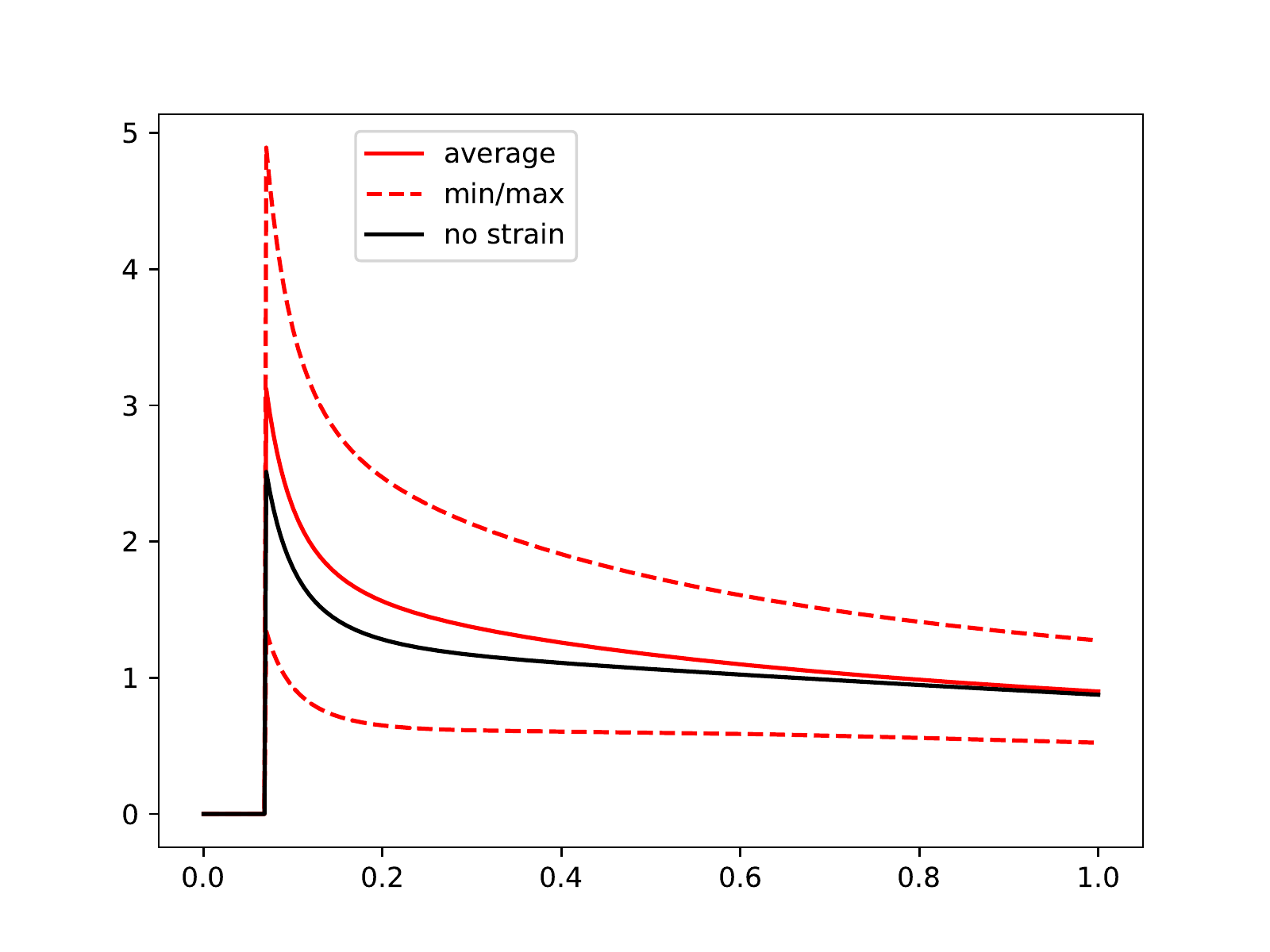}};
\node[inner sep=0pt] (russell) at (1.8,0.95)
    {\includegraphics[width=0.18\textwidth]{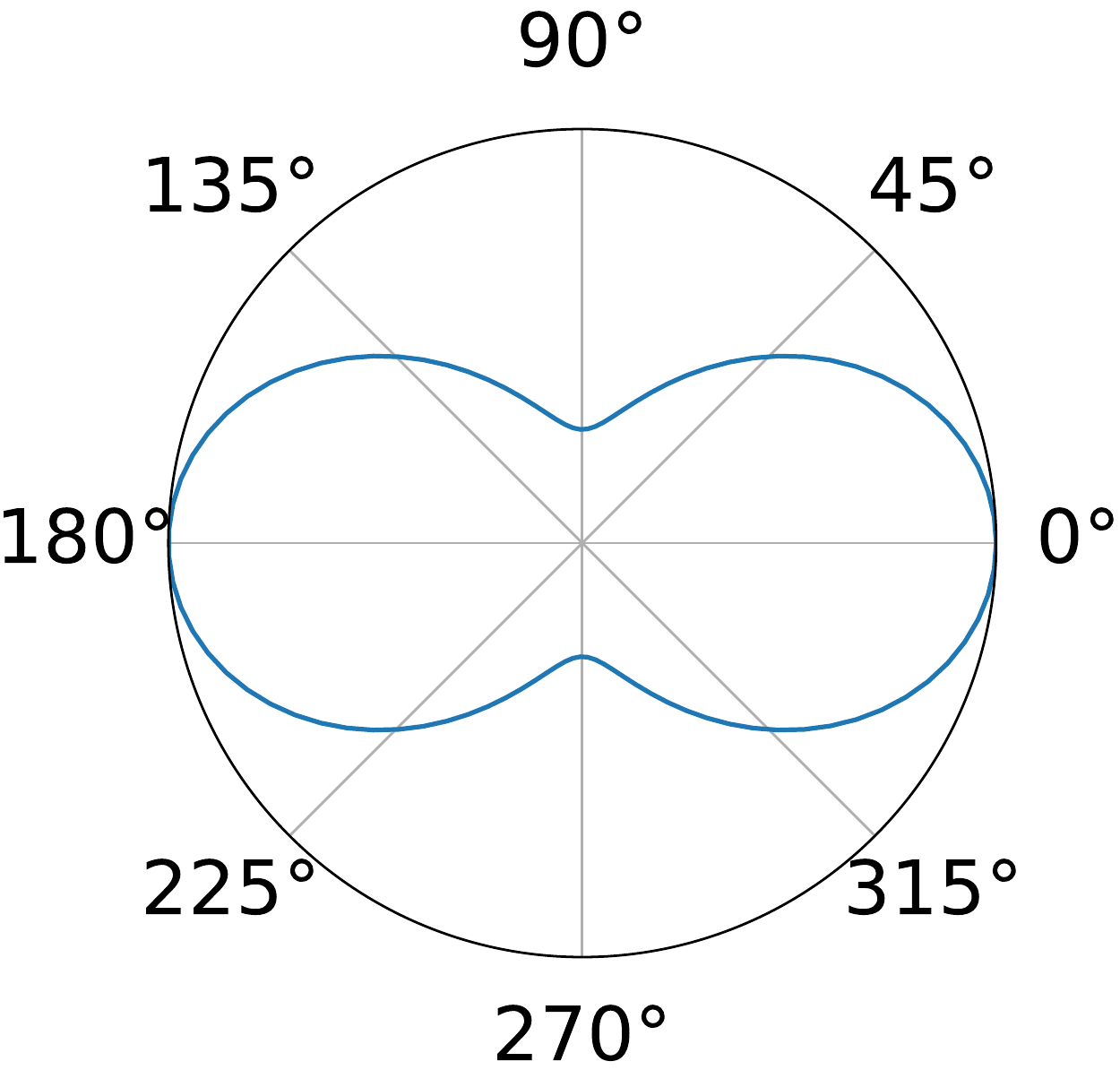}};
\node at (0,-3.3)   {$\hbar\omega$  $(\si{\electronvolt})$};
\node[rotate=90]at (-4.2,0)  {$\mathcal P$ (\%)};
\node at (3,-0.9) {$\theta_A=0$};
\node at (3,-2.15) {$\theta_A=\frac{\pi}{2}$};
\node[] at (1.95,1.7)  {$\mathcal P(\theta_A)$};
\node at (-1.2,1) {4 QL};
\end{tikzpicture}
\caption{Absorbance spectrum of a $4$~QL slab with $\epsilon_{zy}=10\%$ and all other strain components set to zero. The absorbance varies sinusoidally with the polarization angle; here the average and extrema of the absorbance as a function of polarization are shown. The inset shows the band edge absorbance as a function of the polarization angle. The absorbance is either increased or decreased depending on the polarization of the light. However, the average absorbance is increased. Here we have used parameters of the unstrained model from Ref. \onlinecite{njp_films} and $X_i=Y_i=\SI{10}{\electronvolt\angstrom}$. We point to that contributions from QWS to the absorbance set in above $0.7$~eV but these are not included in the plot.}
\label{absorbancewithstrain}
\end{figure}
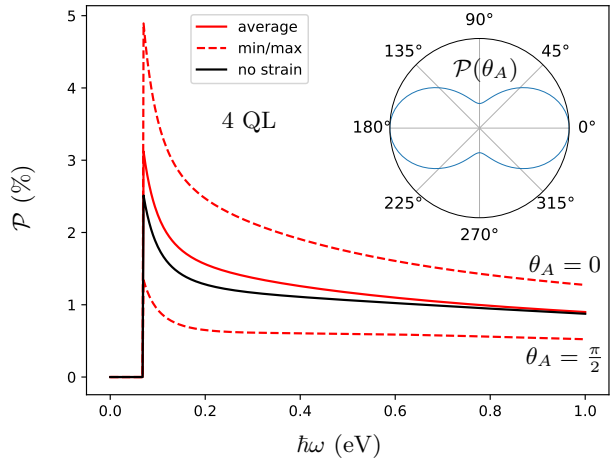

For the more general problem of a slab with shear strain or $\epsilon_{xx}-\epsilon_{yy}\neq 0$ we must integrate Eq.~\eqref{absorbance_general} numerically. Since the isotropy is broken, the absorbance now depends sinusoidally on the polarization angle. Yet, an increase in the absorbance is found when averaged over all polarizations as shown in Fig.~\ref{absorbancewithstrain} for a $4$~QL slab with $\epsilon_{zy}=10\%$, and all other strain components set to zero. Note also the high absorbance that continues to larger photon energies (up to and above $1$~eV) demonstrating the potential of bismuth selenide as an effective photovoltaic and light harvesting material~\cite{Bernardi2013,Sun2014,Sharma_nanowires}.

\begin{figure}
\begin{tikzpicture}
\node[inner sep=0pt] (russell) at (0,0)
    {\includegraphics[width=0.5\textwidth]{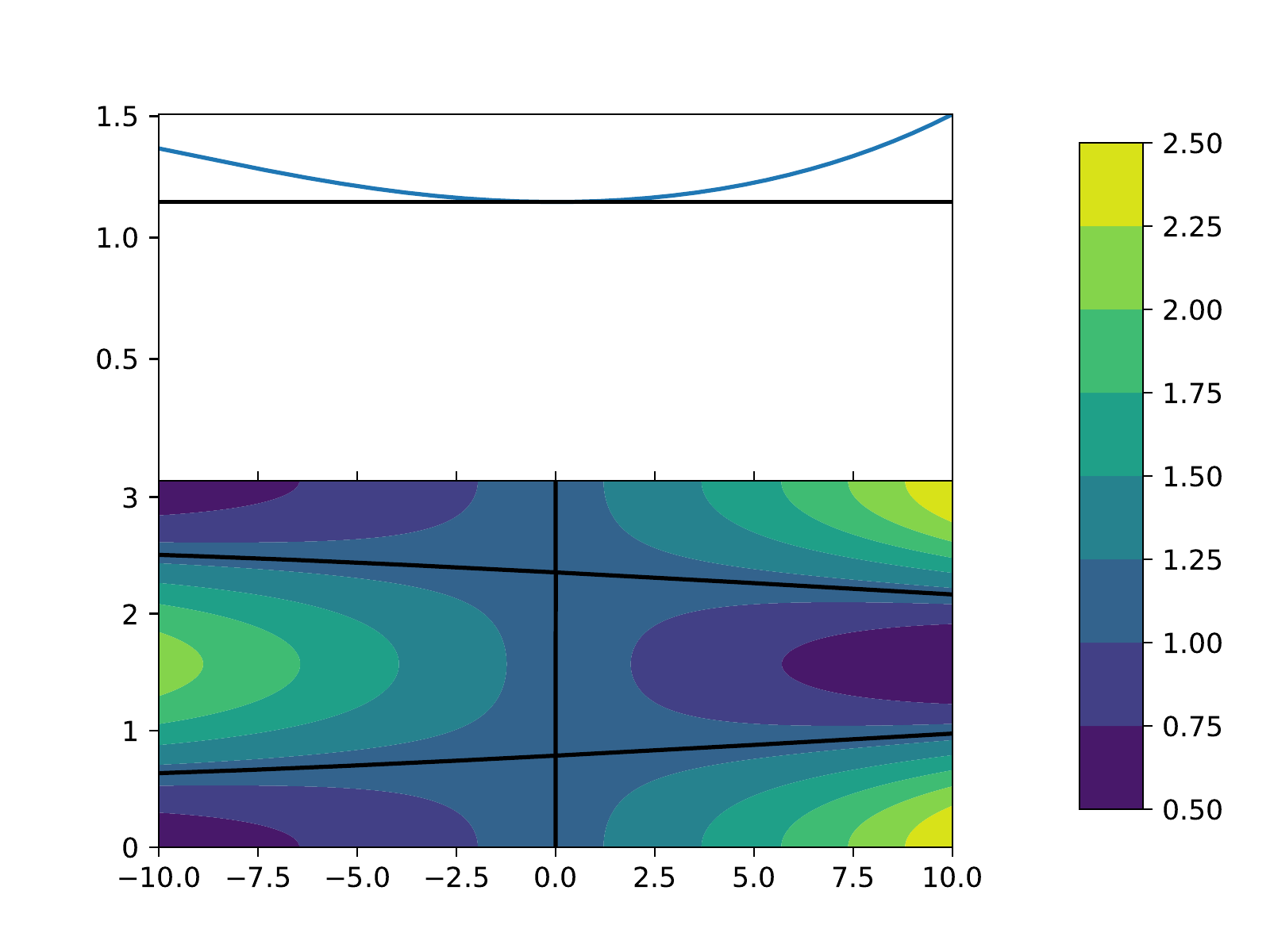}};
    \node at (-0.3,-3.2) {$\epsilon_{zy}$ (\%)};
    \node[rotate=90] at (-4.1,-1.3) {$\theta_A$};
    \node[rotate=90] at (-4.1,1.4) {$\mathcal P_\text{av}$ (\%)};
    \node[rotate=90] at (4.5,0.3) {$\mathcal P$ (\%)};
    \node at (2.5,1.92) {$\frac{\alpha\pi}{2}$};
\end{tikzpicture}
\caption{In the $L\rightarrow \infty$ limit the surface state gap is closed and the absorbance is independent of the photon energy below the bulk band gap. Without strain the absorbance is $\frac{\alpha\pi}{2}\approx1.1\%$ for all polarizations which is exactly half the absorbance of graphene. For slabs with $6$ or more QL's, the gap is too small to be measured and these {\it thick} films are therefore well described by this limit. Due to the anisotropy imposed by shear strain, the absorbance varies with the polarization angle as shown in the contour plot. The black contours show an absorbance of $\frac{\alpha\pi}{2}$. The top plot shows the absorbance averaged over polarizations, showing that the absorbance is always increased by application of shear strain. This increase is a general result independent of model parameters and the type of strain applied as long as the rotational symmetry in the $xy$ plane is broken. Here we have used the Fermi velocity for a $6$~QL slab (without strain) from Ref.~\cite{njp_films} and $X_i=Y_i=\SI{10}{\electronvolt\angstrom}$.}
\label{2Dplot}
\end{figure}
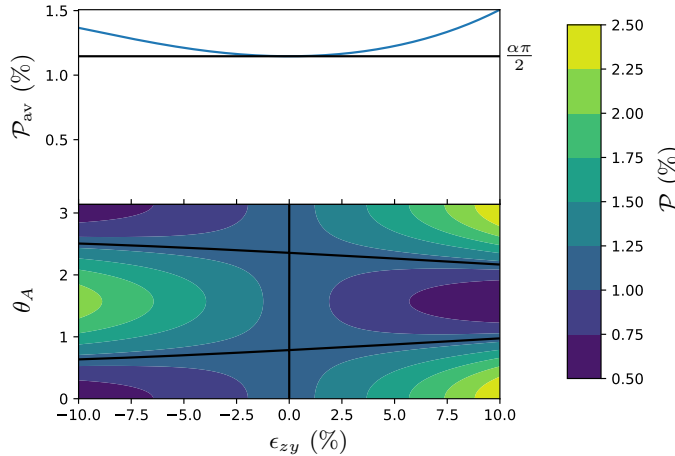

Finally, in the limit of a thick film the absorbance becomes independent of photon energy for an unstrained slab with a universal value given by the fine structure constant $\frac{\alpha\pi}{2}$ independent of polarization which is half the absorbance found for graphene. Including strain the absorbance remains independent of the photon energy, but now we get a sinusoidal dependence on the polarization angle as seen in Fig.~\ref{2Dplot}. If we average over polarizations we can integrate Eq. \eqref{absorbance_general} analytically:
\begin{align}
\mathcal P_\text{av} &= \frac{ \alpha}{4\pi }\int_0^{2\pi} \mathrm d \theta_A \int_0^{2\pi} \mathrm d\theta_k   \frac{ \Im[\chi]^2+|\gamma|^2 }{ A(\theta_k)^4 }.
\end{align}
The integral over $\theta_A$ is straightforward, and using the integral $\int_0^{2\pi}\mathrm d \theta_k \left(a\sin(2\theta_k + \varphi) + b \right)^{-2} = 2\pi b(b^2-a^2)^{-3/2}$ the average absorbance, normalized by $\frac{\alpha\pi}{4}$, is found to be
\begin{multline}
\mathcal P_\text{av}/\left( \frac{\alpha\pi}{4}\right) = \\
\frac{  |\bar\alpha_x|^2 + |\bar \beta_x|^2 + |\bar \alpha_y|^2 + |\bar \beta_y|^2}{\sqrt{ \left( |\bar \alpha_x|^2 + |\bar \beta_x|^2\right)\left( |\bar \alpha_y|^2 + |\bar \beta_y|^2\right) - \Re[\bar \alpha_x \bar \alpha_y^* + \bar \beta_x \bar \beta_y^*]^2 } }
\end{multline}
\begin{align*}
&\geq
 \frac{  |\bar\alpha_x|^2 + |\bar \beta_x|^2 + |\bar \alpha_y|^2 + |\bar \beta_y|^2}{\sqrt{ \left( |\bar \alpha_x|^2 + |\bar \beta_x|^2\right)\left( |\bar \alpha_y|^2 + |\bar \beta_y|^2\right) } }\\
&=\left(\frac{ |\bar\alpha_x|^2 + |\bar \beta_x|^2}{ |\bar\alpha_y|^2 + |\bar \beta_y|^2}\right)^{1/2} +
\left(\frac{ |\bar\alpha_x|^2 + |\bar \beta_x|^2}{ |\bar\alpha_y|^2 + |\bar \beta_y|^2}\right)^{-1/2} \geq 2,
\end{align*}
we arrive at the remarkable result that, independent of the model parameters, the average absorbance can only be increased by application of strain.

In summary, we predict a marked enhancement of the optical absorbance in strained films of Bi$_2$Se$_3$ class TI materials. Without breaking inversion symmetry, we have demonstrated that uniaxial strain can lead to a Van Hove singularity in the JDOS, causing a diverging optical absorbance. Shear strain was shown to inflict a strong dependence on polarization angle in the optical absorbance, and independent of material parameters, the angular averaged absorbance was found always to increase with strain. These findings have strong potential for optimizing the performance of Bi$_2$Se$_3$ TI optoelectronic devices and particularly in photodetector and photovoltaic broadband applications. Tuning the band gap with strain, instead of varying the TI slab thickness, makes it possible to tune a single device for high photodetection performance in a large photon energy range. The application of strain may also be used to enhance the anisotropic photocurrent response reported in Ref.~\onlinecite{Yao2015}. It is shown that the combination of strain and $\epsilon \mathbf k$ terms in the Hamiltonian is crucial for understanding the promising optical absorption properties of Bi$_2$Se$_3$ topological insulators.

MRB and MW gratefully acknowledge financial support from the Danish Council of Independent Research (Natural Sciences) grant no.: DFF-4181-00182. AML gratefully acknowledges financial support from the Carlsberg Foundation.  The Center for Quantum Devices is funded by the Danish National Research Foundation.

\bibliography{references}{}
\bibliographystyle{apsrev4-1}

\end{document}